\documentclass[twocolumn,showpacs,preprintnumbers,amsmath,amssymb]{revtex4}
\usepackage{amssymb}
\usepackage{graphicx}
\usepackage{dcolumn}
\usepackage{bm}
\usepackage{textcomp}

\newtheorem{lemma}{Lemma}

\newtheorem{theorem}{Theorem}

\newcommand\oprod[2]{\ensuremath{|#1\rangle\langle#2|}}

\begin{document}
\title{Improved statistical fluctuation analysis for measurement-device-independent quantum key distribution with three-intensity decoy-state method}
\author{Zong-Wen Yu$ ^{1,2}$, Yi-Heng Zhou$ ^{1,3}$,
and Xiang-Bin Wang$ ^{1,3,4\footnote{Email
Address: xbwang@mail.tsinghua.edu.cn}\footnote{Also a member of Center for Atomic and Molecular Nanosciences at Tsinghua University}}$}

\affiliation{ \centerline{$^{1}$State Key Laboratory of Low
Dimensional Quantum Physics, Department of Physics,} \centerline{Tsinghua University, Beijing 100084,
People¡¯s Republic of China}\centerline{$^{2}$Data Communication Science and Technology Research Institute, Beijing 100191, China}\centerline{$^{3}$ Synergetic Innovation Center of Quantum Information and Quantum Physics, University of Science and Technology of China}\centerline{  Hefei, Anhui 230026, China
 }\centerline{$^{4}$ Shandong
Academy of Information and Communication Technology, Jinan 250101,
People¡¯s Republic of China}}

\begin{abstract}
We present an improved statistical fluctuation analysis for measurement device independent quantum key distribution with three-intensity decoy-state method. Taking the statistical fluctuations for different sources jointly, we present more tightened formulas for some key quantities used in calculating the secure final key. Numerical simulation shows that, given the total number of pulses $10^{12}$, our method  improves the key rate by about 97\%  for a distance of 50kms compared with the result given by Xu., et al. (Phys. Rev. A 89, 052333); and improves the key rate by $146\%$ for a distance of 100kms compared with the result from full optimization of all parameters but treating the statistical fluctuations traditionally, i.e., treating the fluctuations for different sources separately.
\end{abstract}


\maketitle


\section{Introduction}\label{SecIntro}
Quantum key distribution (QKD) is one of the most successful applications of quantum information processing. QKD can provide unconditional security based on the laws of quantum physics~\cite{BB84,GRTZ02}. However, due to the imperfections in real-life implementations of QKD, a large gap between its theory and practice remains unfilled. Security for real set-ups of QKD~\cite{BB84,GRTZ02} has become a major problem in this area. The major imperfectitons in practical QKD are imperfect single-photon source and the limited efficiency of the detectors. Fortunately, by using the decoy-state method~\cite{ILM,H03,wang05,wang06,LMC05,AYKI,haya,peng,wangyang,rep,njp}, it has been shown that the unconditional security of QKD can still be assured with an imperfect single-photon source~\cite{PNS1,PNS}. The limited detection efficiency is another threaten to the security~\cite{lyderson}.
To patch up this, several approaches have been proposed, including the so called device independent QKD (DI-QKD)~\cite{ind1} and the measurement-device independent QKD (MDI-QKD) which was based on the idea of entanglement swapping~\cite{ind3,ind2}. The key idea of MDI-QKD is that both legitimate users, Alice and Bob, are senders. Neither Alice nor Bob performs any measurement, they only send out quantum signals to the un-trusted third party (UTP), who is supposed to perform a Bell state measurement to each pulse pairs. After Alice and Bob send out signals, they wait for UTP's announcement of weather he has obtained a successful event after detection, and proceed to the standard postprocessing of their sifted data.  By using the decoy-state method, Alice and Bob can use imperfect single-photon sources~\cite{ind2,wangPRA2013,we} securely in the MDI-QKD. Hence, the decoy-state MDI-QKD can remove all detector side-channel attacks with imperfect single-photon sources. Because of these important advantages, the decoy-state MDI-QKD has been studied extensively both experimentally~\cite{tittel1,tittel2,liuyang} and theoretically~\cite{wangPRA2013,MaPRA2012,LiangPRA2013,le,wangArxiv,lopa,Wang3int, curtty,Wang3improve,Wang3g,Wang1312, WangModel,Xu1406}.

As is well known, in any real experiment, the key size is finite and we have to consider the effect of statistical fluctuations caused by a finite key size. Such an analysis is crucial to ensure the security of MDI-QKD hence  it has drawn much attentions~\cite{curtty,MaPRA2012,LiangPRA2013,le,Xu1406}. Though a non-zero secure key with statistical fluctuations is proven, the key rate value is rather limited. In all these works, the statistical fluctuations of each sources are considered separately. Relations among the statistical fluctuation of different sources are not considered. Actually, as shown in this paper, the relationship among statistical fluctuations of different sources takes an important role in the key rate improvement. By considering them jointly,  we obtain more tightened bounds of $s_{11}^{Z}$ and $e_{11}^{X}$ which lead to a much higher key rate.

In what follows, we shall first review the decoy-state MDI-QKD for both asymptotic results and non-asymptotic results, and then show our main idea for the improvement through considering the fluctuations of different sources jointly in the first part of section~\ref{sec:Improved}. In the second part of section~\ref{sec:Improved}, we systematically present our improved statistical fluctuation analysis by introducing relations among the fluctuations of different sources. We then present the numerical simulation results in section IV. The article is ended with a concluding remark.

\section{Three-intensity decoy-state method for MDI-QKD}
In the MDI-QKD protocol, each time a pulse-pair (two-pulse state) is sent to the relay for detection. The relay is controlled by an UTP. The UTP will announce whether the pulse-pair has caused a successful event. Those bits corresponding to successful events will be post-selected and further processed for the final key. Since in practice only imperfect single-photon sources are available, we need the decoy-state method for security in practice.

In the three-intensity decoy-state protocol, we assume Alice (Bob) has three different sources, say $o_A$, $x_A$, $y_A$ ($o_B$, $x_B$, $y_B$) which can only emit three different states in photon number space $\rho_{o_A}=\oprod{0}{0}$, $\rho_{x_A}^\omega$, $\rho_{y_A}^\omega$ ($\rho_{o_B}=\oprod{0}{0}$, $\rho_{x_B}^\omega$, $\rho_{y_B}^\omega$) respectively, where the superscript indicates the basis information, $\omega=X$ for $X$ basis and $\omega=Z$ for $Z$ basis.  We denote
\begin{eqnarray}
  \rho_{x_A}^\omega=\sum_{k}a_{k}^\omega\oprod{k}{k}, &\quad& \rho_{y_A}^\omega=\sum_{k}{a_{k}^{\prime \omega}}\oprod{k}{k}, \label{eq:rhoA} \\
  \rho_{x_B}^\omega=\sum_{k}b_{k}^\omega\oprod{k}{k}, &\quad& \rho_{y_B}^\omega=\sum_{k}{b_{k}^{\prime \omega}}\oprod{k}{k}, \label{eq:rhoA}
\end{eqnarray}
where $a_k^{\omega}$, $a_k^{\prime\omega}$, $b_k^{\omega}$ and $b_k^{\prime\omega}$ are nonnegative parameters.
We shall consider the decoy-state method in each basis separately. For simplicity, we shall omit the superscripts $\omega$  in what follows of this article provided that the omission does not cause any confusion. In particular, since we assume to implement the decoy-state method in each bases, when we say
any source $lr$, we always mean source $lr$ in a certain basis $\omega$.
We request the states above satisfy the following very important condition
\begin{equation}\label{eq:cond}
  \frac{a_k^{\prime}}{a_k}\geq \frac{a_2^{\prime}}{a_2}\geq \frac{a_1^{\prime}}{a_1}, \quad
  \frac{b_k^{\prime}}{b_k}\geq \frac{b_2^{\prime}}{b_2}\geq \frac{b_1^{\prime}}{b_1},
\end{equation}
for $k\geq 2$ so that the decoy-state results can apply\cite{wangPRA2013}. Imperfect sources used in practice such as the coherent state source, the heralded source out of the parametric-down conversion, satisfy the above conditions.

At each time, Alice (Bob) randomly chooses source ${l_A}$ ($r_B$) with probability $p_{l_A}$ ($p_{r_B}$). Conditional on this source selection, she (he) chooses basis $\omega$  with probability $p_{\omega|l_A}$ ($p_{\omega|r_B}$). Here $l=o,x,y$ ($r=o,x,y$). The pulse from Alice and the pulse from Bob form a pulse pair and are sent to UTP. We regard equivalently that each time a two-pulse source is selected and a pulse pair (one pulse from Alice, one pulse from Bob) is emitted. For postprocessing, Alice and Bob evaluate the data sent in two bases separately. The $Z$-basis is used for key generations, while the $X$-basis is used for testing against tampering and the purpose of quantifying the amount of privacy amplification needed. Here, we use the capital letters $Z(X)$ for the bases and the lowercase letters $o,x,y$ for the different sources. Here and after, we omit the subscripts $A$ and $B$ provided that this does not cause any confusions.
\subsection{Asymptotic case}\label{subsec:Asymptotic}
As shown in Ref.\cite{wangPRA2013}, we denote $lr$ as the two-pulse source when Alice uses source $l$ and Bob uses $r$, and $l,r$ can take $o,x,y $. For example, two-pulse source $oy$ denote the case when Alice use vacuum source $o$ and Bob uses the signal source $y$.  There are nine two-pulse sources $lr$ in each bases of the three-intensity protocol. We also denote $S_{lr}$ as the yield of two-pulse source $lr$ (in a certain basis, $X$ or $Z$).  $S_{lr}$ are observed values and will be regarded as known values here. However, the yields $s_{mn}^{lr}$ for the two-pulses states $|m\rangle\langle m| \otimes |n\rangle\langle n|$ out of source $lr$  cannot be directly observed.
In the asymptotic case,  we assume that $s_{mn}^{lr}$ for all $lr$ are the same and we can denote all of them by $s_{mn}$, i.e.
\begin{equation}\label{id}
s_{mn}^{lr}=s_{mn},
\end{equation}
for all $l,r=o,x,y$. Given this, we can formulate the very important unknown variable $s_{11}$ by using relations
\begin{equation}\label{cons}
S_{lr} = \sum_{m,n} c^{lr}_{mn}s_{mn},
\end{equation}
if the state for the two-pulse source $lr$ is
\begin{equation}
\rho_{lr}= \sum_{m,n} c^{lr}_{mn}|m\rangle \langle m|\otimes |n\rangle \langle n|.
\end{equation}

In order to calculate the secret final key rate of this protocol, we need the lower bound of the yield $s_{11}$ and the upper bound of the error rate $e_{11}$.
In Ref.~\cite{wangPRA2013}, Wang presented the first explicit formula for the practical decoy-state implementation through using part of the above constraints given by Eq.(\ref{cons}), sources $lr(l,r=o,x,y)$ except $xy$ and $yx$. Without losing the generality, we assume $K_a=\frac{a_1^{\prime}b_2^{\prime}}{a_1 b_2}\leq \frac{a_2^{\prime}b_1^{\prime}}{a_2 b_1}=K_b$. Then the lower bound of $s_{11}$ can be estimated by the following explicit formula~\cite{wangPRA2013}
\begin{equation}\label{eq:s11of14}
  \underline{\hat{s}}_{11}^{(1)}=\frac{\hat{S}_{+}^{(1)}- \hat{S}_{-}^{(1)}}{a_1 a_1^{\prime} \tilde{b}_{12}},
\end{equation}
where $\tilde{b}_{12}=b_1 b_2^{\prime}-b_1^{\prime} b_2$, $\hat{S}_{+}^{(1)}=\mathcal{S}_{+}^{(1)}(S_{xx},S_{oy},S_{yo},S_{oo})$ and $\hat{S}_{-}^{(1)}=\mathcal{S}_{-}^{(1)}(S_{yy},S_{ox},S_{xo})$ with functions
\begin{eqnarray}
  \mathcal{S}_{+}^{(1)}(z_1,z_2,z_3,z_4)&=& a_{1}^{\prime} b_2^{\prime} z_1 +a_1 b_2 a_0^{\prime} z_2 +a_1 b_2 b_0^{\prime} z_3 \nonumber \\
  & & +(a_1^{\prime} b_2^{\prime} a_0 b_0 -a_1 b_2 a_0^{\prime} b_0^{\prime}) z_4,  \label{eq:DefSp1} \\
  \mathcal{S}_{-}^{(1)}(z_1,z_2,z_3)&=& a_1 b_2 z_1 +a_1^{\prime} b_2^{\prime} a_0 z_2 +a_1^{\prime} b_2^{\prime} b_0 z_3. \label{eq:DefSn1}
\end{eqnarray}
Furthermore, in the case of $K_a>K_b$, the lower bound of $s_{11}$ can be calculated with Eq.(\ref{eq:s11of14}) by making the exchange between $a_k$ and $b_k$, and the exchange between $a_k^{\prime}$ and $b_k^{\prime}$ for $k=1,2$.

Besides this formula, we also present another formula to estimate the lower bound of $s_{11}$ for this three-intensity protocol~\cite{Wang3int}, through using another part of constraints given by Eq.(\ref{cons}), i.e., the constraints for sources $lr(l,  r=o,x,y)$ except $yy$. Explicitly, the lower bound of $s_{11}$ can be estimated by the following explicit formula
\begin{equation}\label{eq:s11of123}
  \underline{\hat{s}}_{11}^{(2)}=\frac{\hat{S}_{+}^{(2)}-\hat{S}_{-}^{(2)}}{a_1 b_1 \tilde{a}_{12} \tilde{b}_{12}},
\end{equation}
where $\tilde{a}_{12}=a_1 a_2^{\prime}-a_1^{\prime} a_2$, $\tilde{b}_{12}=b_1 b_2^{\prime} -b_1^{\prime} b_2$, $\hat{S}_{+}^{(2)}=\mathcal{S}_{+}^{(2)}(S_{xx},S_{oy},S_{yo},S_{oo})$ and $\hat{S}_{-}^{(2)}=\mathcal{S}_{-}^{(2)}(S_{xy},S_{yx},S_{ox},S_{xo})$ with
\begin{eqnarray}
  \mathcal{S}_{+}^{(2)}(z_1,z_2,z_3,z_4)= g_{xx}z_1 +g_{oy}z_2 +g_{yo}z_3 +g_{oo}z_4, \label{eq:s11of123Sp} \\
  \mathcal{S}_{-}^{(2)}(z_1,z_2,z_3,z_4)= g_{xy}z_1 +g_{yx}z_2 +g_{ox}z_3 +g_{xo}z_4, \label{eq:s11of123Sn}
\end{eqnarray}
and
\begin{eqnarray*}
  g_{xx}&=& a_1 a_2^{\prime} b_1 b_2^{\prime}- a_1^{\prime} a_2 b_1^{\prime} b_2, \quad g_{xy}\;=\; b_1 b_2\tilde{a}_{12}, \\
  g_{yx}&=& a_1 a_2 \tilde{b}_{12}, \quad g_{oy}\;=\; a_0 g_{xy}, \quad g_{yo}\;=\; b_0 g_{yx}, \\
  g_{oo}&=& a_0 b_0 g_{xx}-a_0 b_0^{\prime} g_{xy} -a_0^{\prime} b_0 g_{yx} \\
  &=& a_0 b_2 \tilde{a}_{12}\tilde{b}_{01} + b_0 a_1 \tilde{a}_{02}\tilde{b}_{12}, \\
  g_{ox}&=& a_0 g_{xx} -a_0^{\prime} g_{yx} =a_1 \tilde{a}_{02} \tilde{b}_{12}+ a_0 b_1^{\prime} b_2 \tilde{a}_{12}, \\
  g_{xo}&=& b_0 g_{xx} -b_0^{\prime} g_{xy} =b_0 a_1 a_2^{\prime} \tilde{b}_{12} +b_2 \tilde{a}_{1} \tilde{b}_{01}.
\end{eqnarray*}
In above equations, we denote $\tilde{a}_{02}=a_0 a_2^{\prime}-a_0^{\prime} a_2$ and $\tilde{b}_{01}=b_0 b_1^{\prime} -b_0^{\prime} b_1$. With the conditions listed in Eq.(\ref{eq:cond}), we can easily prove that $g_{lr}\geq 0$ for all $l,r=o,x,y$.

As discussed in Ref.~\cite{Wang3int}, we know that the lower bound $\underline{\hat{s}}_{11}^{(2)}$ is always better than $\underline{\hat{s}}_{11}^{(1)}$ in the asymptotic case. Whereas, in the non-asymptotic case, we need reanalysis the relation between them. Actually, the priority of $\underline{\hat{s}}_{11}^{(2)}$ will disappear in the case of reasonable data-size for a long enough key distribution distance.

Besides the lower bound of $s_{11}$, we can estimate the upper bound of $e_{11}$ with the following explicit formula
\begin{equation}\label{eq:e11UX}
  \hat{\overline{e}}_{11}=(\hat{T}_{+} -\hat{T}_{-})/(a_1 b_1 \underline{\hat{s}}_{11}),
\end{equation}
where
\begin{equation}\label{eq:e11UXpn}
  \hat{T}_{+}=T_{xx}+a_0 b_0 T_{oo}, \quad \hat{T}_{-}=a_0 T_{ox}+b_0 T_{xo},
\end{equation}
and $\underline{\hat{s}}_{11}$ is the lower bound of $s_{11}$ which can be estimated by using Eq.(\ref{eq:s11of123}).

\subsection{Non-asymptotic case}\label{subsec:Nonasymptotic}
In any real experiment, the total pulses sent by Alice and Bob are finite.
So the number of sifted keys is always finite. In order to extract the secure final key,
we have to consider the effect of statistical fluctuations caused by the finite-size key.
In the non-asymptotic case, yields of the same two-pulse state out of different sources are not always equal to each other rigorously.
That is to say, we need treat them {\em differently}, i.e.,
\begin{equation}\label{eq:smnlr}
s_{mn}^{lr}\neq   {s_{mn}^{l'r'}},
\end{equation}
for different two-pulse sources $lr$ and $l^{\prime}r^{\prime}$ $(lr\neq l^{\prime}r^{\prime})$. For example, for two-pulse sources $xx$ and $xy$, we have $s_{mn}^{xx}\neq s_{mn}^{xy}$ from Eq.(\ref{eq:smnlr}) with $lr=xx$ and $l^{\prime}r^{\prime}=xy$. In such a case, there are too many  variants $\{s_{mn}^{lr}|m,n\geq 0;l,r=o,x,y\}$. To obtain the lower bound value for $s_{11}$ and the upper bound value for $e_{11}$, one can implement the idea of Ref.\cite{njp}, i.e., treating the averaged yield of a specific state from different  sources. As was shown there~\cite{njp}, in the BB84 decoy-state method, one can introduce the averaged value for the yield of an $m$-photon state from all sources in the same basis. The same idea can obviously apply for the decoy-state method MDI-QKD, i.e., treat these variants $s_{mn}^{lr}$ uniformly by introducing the mean values. Accordingly, define $\langle{s}_{mn}\rangle$ as the mean value of yield of state $|mn\rangle$ produced by all sources used in the decoy-state method (in a certain basis),
\begin{equation}
  \langle{s}_{mn}\rangle = \sum_{{lr}} p_lp_r c_{mn}^{lr} s_{mn}^{lr}.
\end{equation}
Based on this, we can also {\em define} quantity
\begin{equation}\label{cons1}
  \langle{S}_{lr}\rangle= \sum_{m,n=0}^\infty c_{mn}^{lr}\langle{s}_{mn}\rangle.
\end{equation}
Replacing $S_{lr}$ by $\langle{S}_{lr}\rangle $ in Eq.(\ref{cons}), we can formulate the lower bound of $\langle{s}_{11} \rangle $.
Note that even though $S_{lr}$ are known values directly observed in an experiment, $\langle{S}_{lr}\rangle$  are not.
However, given the values $S_{lr}$ and $N_{lr}$, we have
\begin{equation}\label{eq:ExptS}
 \langle{S}_{lr}\rangle = S_{lr}\left(1+  \delta_{lr}\right).
\end{equation}
With a probability larger than $1-\epsilon$, $\delta_{lr}$ is in the range of
\begin{equation}\label{eq:defDelta}
|\delta_{lr}|\le n_\delta \sqrt{\frac{1}{{N}_{lr} S_{lr}  }}\triangleq \overline{\delta}_{lr},
\end{equation}
where $N_{lr}$ is the number of pulses sent out by Alice and Bob when they use sources ${l}$ and ${r}$ respectively, $n_{\delta}$ is the number of standard deviations one chooses for statistical fluctuation analysis with the given security bound. With these notations, we know that $N_{lr}S_{lr}$ is the number of successful event announced by UTP when Alice and Bob use sources ${l}$ and ${r}$ respectively.

In Ref.\cite{njp}, $n_\delta$ is set to be 10. Here in this paper we shall set
\begin{equation}\label{poss2}
  n_\delta=5.3,
\end{equation}
which corresponds to $\epsilon = 10^{-7}$\cite{curtty,Xu1406}
in our numerical simulation, so as to make a fair comparison with \cite{Xu1406}.
Therefore,  we can formulate the lower bound value of $\langle{s}_{11}\rangle$ by $\langle{S}_{lr}\rangle$.

In order to get a reasonable lower bound of $s_{11}$ in the non-asymptotic case, we reconsider the explicit formulas Eq.(\ref{eq:s11of14}) first. As discussed above, in the non-asymptotic case, the observed values are different from its mean values. So we need to replace $S_{lr}$ by its mean values $\langle{S}_{lr}\rangle$ defined in Eq.(\ref{eq:ExptS}). Then the formula turns into a function of quantities  $\delta_{lr}$.

In the security proof, we assume that Eve can do anything except to violate rules of the nature. In order to obtain a reasonable estimation of the lower bound of $s_{11}$, we should find out the worst case under the constraints about $\delta_{lr}$ given by Eq.(\ref{eq:defDelta}). If one simply treats all $\delta_{lr}$ separately,  the worst case result is
\begin{equation}\label{eq:s11of14sta}
  \underline{s}_{11}^{(1)}=\frac{\underline{S}_{+}^{(1)}-\overline{S}_{-}^{(1)}}{a_1 a_1^{\prime}\tilde{b}_{12}},
\end{equation}
where $\underline{S}_{+}^{(1)}=\mathcal{S}_{+}^{(1)}(\underline{S}_{xx}, \underline{S}_{oy}, \underline{S}_{yo}, \underline{S}_{oo})$, $\overline{S}_{-}^{(1)}=\mathcal{S}_{-}^{(1)}(\overline{S}_{yy},\overline{S}_{ox}, \overline{S}_{xo})$ with $\mathcal{S}_{+}^{(1)}$, $\mathcal{S}_{-}^{(1)}$ being defined in Eqs.(\ref{eq:DefSp1},\ref{eq:DefSn1}) respectively, and
\begin{equation}\label{eq:SlrLU}
  \underline{S}_{lr}=S_{lr}(1-\overline{\delta}_{lr}), \quad
  \overline{S}_{lr}=S_{lr}(1+\overline{\delta}_{lr}).
\end{equation}
for all $l,r=o,x,y$. In above equations, $\overline{\delta}_{lr}$ is the upper bound of the  $\delta_{lr}$ given by Eq.(\ref{eq:defDelta}).

Besides this lower bound, we can also obtain the other one from Eq.(\ref{eq:s11of123}) in the same way. Explicitly, we have
\begin{equation}\label{eq:s11of123sta}
  \underline{s}_{11}^{(2)}=\frac{\underline{S}_{+}^{(2)}-\overline{S}_{-}^{(2)}}{a_1 b_1 \tilde{a}_{12}\tilde{b}_{12}},
\end{equation}
where $\underline{S}_{+}^{(2)}=\mathcal{S}_{+}^{(2)}(\underline{S}_{xx}, \underline{S}_{oy}, \underline{S}_{yo}, \underline{S}_{oo})$, $\overline{S}_{-}^{(2)}=\mathcal{S}_{-}^{(2)}(\overline{S}_{xy},\overline{S}_{yx}, \overline{S}_{ox}, \overline{S}_{xo})$ with $\mathcal{S}_{+}^{(2)}$, $\mathcal{S}_{-}^{(2)}$ being defined in Eqs.(\ref{eq:s11of123Sp},\ref{eq:s11of123Sn}) respectively, $\underline{S}_{lr}$ and $\overline{S}_{lr}$ being defined in Eq.(\ref{eq:SlrLU}).

In Ref.~\cite{Wang3int}, we have shown that the lower bound $\underline{\hat{s}}_{11}^{(2)}$ is always better than $\underline{\hat{s}}_{11}^{(1)}$ with the same experimental parameters in the asymptotic case. However, in the non-asymptotic case, the lower bound $\underline{s}_{11}^{(1)}$ can be better than $\underline{s}_{11}^{(2)}$ in the case of reasonable data-size for a long enough key distribution distance. So we should choose the bigger one. Explicitly, we define the new lower bound of $s_{11}$ for this three-intensity protocol as follows
\begin{equation}\label{eq:s11sta}
  \underline{s}_{11}=\max\{\underline{s}_{11}^{(1)},\underline{s}_{11}^{(2)}\},
\end{equation}
where $\underline{s}_{11}^{(1)}$ and $\underline{s}_{11}^{(2)}$ are defined in Eq.(\ref{eq:s11of14sta}) and Eq.(\ref{eq:s11of123sta}) respectively.

Similarly, one can also work out the averaged value of $\langle{e}_{11}\rangle$ by $\langle{T}_{lr}\rangle$ with \begin{equation}\label{eq:DefHatT}
  \langle{T}_{lr}\rangle=\sum_{m,n=0}^\infty c_{mn}^{lr}\langle{s}_{mn}\rangle\langle{e}_{mn}\rangle,
\end{equation}
being the error yields. Here in Eq.(\ref{eq:DefHatT}), we define the mean value $\langle{s}_{mn}\rangle\langle{e}_{mn}\rangle=\sum_{lr} p_l p_r c_{mn}^{lr} s_{mn}^{lr} e_{mn}^{lr}$.
By introducing the relative fluctuations  $\tau_{lr}$, we can write the relation between error yields $\langle{T}_{lr}\rangle$ and the observed value $T_{lr}$ as follows
\begin{equation}\label{eq:ExptT}
  \langle{T}_{lr}\rangle=T_{lr}(1+\tau_{lr}),
\end{equation}
for all $l,r=o,x,y$. Similarly, after giving the security bound,  $\tau_{lr}$ can be bounded by
\begin{equation}\label{eq:defTau}
  |\tau_{lr}|\leq n_{\tau}\sqrt{\frac{1}{N_{lr}T_{lr}}}\triangleq \overline{\tau}_{lr},
\end{equation}
where $n_{\tau}$ is the number of standard deviations one chooses for statistical fluctuation analysis with the given security bound, $N_{lr}T_{lr}$ is the error count when Alice and Bob use sources ${l}$ and ${r}$ respectively.

In this three-intensity protocol, we can use the following explicit formula to estimate the upper bound of $e_{11}$
\begin{equation}\label{eq:e11sta}
  \overline{e}_{11}=(T_{+}-T_{-})/(a_1 b_1 \underline{s}_{11}),
\end{equation}
where
\begin{equation}\label{eq:e11stapn}
  T_{+}=\overline{T}_{xx}+ a_0 b_0 \overline{T}_{oo}, \quad T_{-}=a_0 \underline{T}_{ox} + b_0 \underline{T}_{xo},
\end{equation}
with $\underline{s}_{11}$ being the lower bound of $s_{11}$ which can be estimated by using Eq.(\ref{eq:s11sta}) and
\begin{equation}\label{eq:TlrLU}
  \underline{T}_{lr}=T_{lr}(1-\overline{\tau}_{lr}), \quad
  \overline{T}_{lr}=T_{lr}(1+\overline{\tau}_{lr}).
\end{equation}
In the above equations, $\overline{\tau}_{lr}$ is the upper bound of  $\tau_{lr}$ defined in Eq.(\ref{eq:defTau}).

Here in this work, instead of using this simple worst-case calculation~\cite{Xu1406}, we propose a  more efficient method to treat the statistical fluctuations in the decoy-state MDI-QKD. In our method, {\em we don't have to consider the fluctuation of each quantities separately.} For example, in estimating the quantity $T_{-}$ in Eq.(\ref{eq:e11stapn}), in a symmetric protocol where $a_0=b_0$, we need to calculate bound of $T_{ox}+T_{xo}$. The simple worst-case result would calculate the worst-case fluctuation for $T_{xo}$ and $T_{ox}$ {\em separately}. However, we can treat this more efficiently by considering the statistical fluctuations {\em jointly}. Say, we regard sources ${ox}$ and ${xo}$ as one source $ox+xo$ which emits state $\frac{1}{2}(\rho_{ox}+\rho_{xo})$. For such a source, the error yield $T_{xo}+T_{ox}=2T_{ox+xo}$. We then only need to consider the fluctuation for only {\em one}  quantity $T_{ox+xo}$. This will improve the performance of the decoy-state protocol. In the next section we present a systematic study of this joint constraints in the statistical fluctuation.

\section{Improved statistical fluctuation analysis}\label{sec:Improved}
In order to estimate the lower bound of $s_{11}$ and the upper bound of $e_{11}$, we need the values of yields $S_{lr}$ and error yields $T_{lr}$ ($l,r=o,x,y$ for this three-intensity decoy-state protocol), which can be observed in experiment. On the other hand, in any real experiment, we have to consider the effect of statistical fluctuation caused by a finite-size key. As discussed above, we need to introduce quantities $\delta_{lr}$ and $\tau_{lr}$ to obtain the values of yields $\langle{S}_{lr}\rangle$ and error yields $\langle{T}_{lr}\rangle$ with its observed values $S_{lr}$ and $T_{lr}$. With a given security bound, we can bound  $\delta_{lr}$ and $\tau_{lr}$, such as the relations presented in Eq.(\ref{eq:defDelta}) and Eq.(\ref{eq:defTau}). In all previous works, the all fluctuations $\delta_{lr}$ for different $lr$ are treated separately and independently and so do all quantities of $\tau_{lr}$ for different $lr$.   In this section, we will introduce some relations among them first. With these relations, we then present improved formulas to estimate the lower bound of $s_{11}$ and the upper bound of $e_{11}$ which lead to a much higher rate in distilling the secure final key.

\subsection{Relations among the fluctuations of different sources}\label{subsec:Relations}
When we do the statistical fluctuation analysis, we need to choose a proper security bound first. With a given definite security bound, we can bound $\delta_{lr}$ and $\tau_{lr}$ by Eq.(\ref{eq:defDelta}) and Eq.(\ref{eq:defTau}) respectively. In order to obtain the relations among these quantities, we need to reconsider the grouping of the successful events announced by UTP.

For clarity, we consider the relation between $\tau_{ox}$ and $\tau_{xo}$ first. As defined above, we know that $\tau_{ox}$ and $\tau_{xo}$ are the relative fluctuations for the observed error yields ${T}_{ox}$ and ${T}_{xo}$ respectively. These two observable are corresponding to the successful events with different two-pulse sources  $ox$ and $xo$. If we group all those successful events of these two sources together, and denote $J=\{ox,xo\}$, the relation between the error yield $T_J$ and its mean value $\langle T_J\rangle$ is
\begin{equation}\label{eq:GainJ12}
  \langle{T}_{J}\rangle= T_{J}(1+\tau_{J}),
\end{equation}
where $\tau_{J}$ is the relative fluctuation for the observable $T_{J}$. Similarly,  $\tau_{J}$ has the following property with given security bound
\begin{equation}\label{eq:defTauJ12}
  |\tau_{J}|\leq \frac{n_{\tau}}{\sqrt{N_{ox}S_{ox} +N_{xo}S_{xo}}}\triangleq \bar{\tau}_{J}.
\end{equation}
In this relation, $N_{ox}S_{ox} +N_{xo}S_{xo}$ is the number of error counts when Alice and Bob use two-pulse sources  $ox$ and $xo$. Reconsidering the definition of $T_{J}$, we know that
\begin{equation*}
  N_{ox}T_{ox} +N_{xo}T_{xo} =(N_{ox} +N_{xo})T_{J}.
\end{equation*}

Now we take into account the mean values. The above relation can be written into
\begin{eqnarray*}
  & &N_{ox}\langle{T}_{ox}\rangle +N_{xo}\langle{T}_{xo}\rangle \\
  &=&N_{ox}T_{ox}(1+\tau_{ox}) +N_{xo}T_{xo}(1+\tau_{xo}) \\
  &=&(N_{ox}T_{ox} +N_{xo}T_{xo}) (1+\tau_{J}).
\end{eqnarray*}
Then we have the relation between these two quantities  $\tau_{ox}$ and $\tau_{xo}$
\begin{eqnarray}\label{eq:defTauoxo}
  |N_{ox} {T}_{ox}\tau_{ox} +N_{xo}{T}_{xo}\tau_{xo}|
  \leq n_{\tau}\sqrt{N_{ox}{T}_{ox} +N_{xo}{T}_{xo}}.
\end{eqnarray}
 Here we have used Eq.(\ref{eq:defTauJ12}). Similarly, the relation between any two different quantities $\tau_{l_1 r_1}$ and $\tau_{l_2 r_2}$ can be written into
\begin{eqnarray}\label{eq:defTauJ2}
  |N_{l_1 r_1}{T}_{l_1 r_1}\tau_{l_1 r_1} +N_{l_2 r_2}{T}_{l_2 r_2}\tau_{l_2 r_2}| \nonumber \\
  \leq n_{\tau}\sqrt{N_{l_1 r_1}{T}_{l_1 r_1} +N_{l_2 r_2}{T}_{l_2 r_2}},
\end{eqnarray}
with $l_1,r_1,l_2,r_2$ each can be any one of $o,x$ or $y$ and $l_1r_1\neq l_2r_2$.

Generally, we can group the successful events of a number of  two-pulse sources together. To see it more clearly, we define the set
\begin{equation}
  {\bf J}=\{lr|l,r=o,x,y\}
\end{equation}
  as the whole set of all possible two-pulse sources (in a certain basis)  used by Alice and Bob in the protocol. Explicitly,
  ${\bf J} = \{oo, ox, xo, oy,yo,xx,xy,yx,yy\}$ in any basis.
With this notation, we can write all the relations among $\tau_{lr}$ into
\begin{equation}\label{eq:defTauAllRelations}
  \left|\sum_{lr\in \mathcal{J}} N_{lr}T_{lr}\tau_{lr}\right| \leq n_{\tau}\sqrt{\sum_{lr\in \mathcal{J}}N_{lr} T_{lr}},
\end{equation}
for all nonempty $\mathcal{J}\subseteq {\bf J}$ and
\begin{equation}\label{eq:defTauNab}
  \sum_{lr\in {\bf J}} N_{lr} T_{lr} \tau_{lr}=0.
\end{equation}
The last equation is deduced from the fact that
\begin{equation*}
  \sum_{lr\in {\bf J}} N_{lr} \langle{T}_{lr}\rangle= \sum_{lr\in {\bf J}} N_{lr} T_{lr}.
\end{equation*}
Specifically, if  set $\mathcal{J}$ contains only one element  $lr$, then the relations presented in Eq.(\ref{eq:defTauAllRelations}) is just the bound for $\tau_{lr}$ shown in Eq.(\ref{eq:defTau}). If set $\mathcal{J}$ contains two elements, say $\mathcal {J}=\{l_1r_1,l_2r_2\}$, then the relations presented in Eq.(\ref{eq:defTauAllRelations}) is just the relation between two quantities $\tau_{l_1 r_1}$ and $\tau_{l_2 r_2}$ given by Eq.(\ref{eq:defTauJ2}).

It should be noted that there are nine different  $\tau_{lr}$ in Eq.(\ref{eq:ExptT}) for this three-intensity protocol in each basis, if we consider all possible $lr$.  Furthermore, in Eq.(\ref{eq:defTauAllRelations}) and Eq.(\ref{eq:defTauNab}), there are $\sum_{k=2}^{9}\mathcal{C}_{9}^{k}=2^9-9-1=502$ joint constraints and $\mathcal{C}_{9}^{1}=9$ boundary constraints for these nine quantities. It is a hard work to obtain an explicit formula to estimate the lower bound of $\langle{t}_{11}\rangle=\langle{s}_{11}\rangle\langle{e}_{11}\rangle$ from Eq.(\ref{eq:DefHatT}) with all these constraints. In the next subsection, we will present some explicit formulas to upper bound $\langle{e}_{11}\rangle$ and lower bound $\langle{s}_{11}\rangle$.

Similarly, all the relations among quantities  $\delta_{lr}$ for all possible $lr$ can be written into
\begin{equation}\label{eq:defDeltaAllRelations}
  \left|\sum_{lr\in \mathcal{J}} N_{lr}S_{lr}\delta_{lr}\right| \leq n_{\delta}\sqrt{\sum_{lr\in \mathcal{J}}N_{lr} S_{lr}},
\end{equation}
for all nonempty $\mathcal{J}\subseteq {\bf{J}}$ and
\begin{equation}\label{eq:defDeltaNab}
  \sum_{lr\in {\bf J}} N_{lr} S_{lr} \delta_{lr}=0.
\end{equation}
Specifically, if set $\mathcal{J}$ contains only one element  $lr$, then the relations presented in Eq.(\ref{eq:defDeltaAllRelations}) is just the bounds for each quantities $\delta_{lr}$ shown in Eq.(\ref{eq:defDelta}).

\subsection{Formulas for improved analysis of statistical fluctuations}\label{subsec:Improved}
Consider the upper bound of $e_{11}$ first. As shown in Eq.(\ref{eq:ExptT}), the mean value $\langle{T}_{lr}\rangle$ is a function of $\tau_{lr}$. Replacing $T_{lr}$ by its mean value $\langle{T}_{lr}\rangle$ in Eq.(\ref{eq:e11UX}), we get a function about quantities $\tau_{lr}$
\begin{equation}\label{eq:e11Xf}
  e_{11}^{f}=(T_{+}^{f}-T_{-}^{f})/(a_1 b_1 \underline{s}_{11}^{f}),
\end{equation}
where
\begin{eqnarray*}
  T_{+}^{f}&=& \hat{T}_{+}+T_{xx}\tau_{xx}+a_0 b_0 T_{oo} \tau_{oo}, \\
  T_{-}^{f}&=& \hat{T}_{-}+a_0 T_{ox}\tau_{ox}+b_0 T_{xo} \tau_{xo},
\end{eqnarray*}
with $\hat{T}_{\pm}$ being the constant factors given by Eq.(\ref{eq:e11UXpn}), $\underline{s}_{11}^{f}$ being the lower bound of $s_{11}$ that will be discussed below.

In order to obtain a proper estimation of the upper bound of $e_{11}$ from the function $e_{11}^{f}$, we need to find out the worst case under the constraints about the quantities $\tau_{lr}$. That is to say, we need maximize the function $e_{11}^{f}$ of variables $\tau_{lr}$ under the constraints shown in Eqs.(\ref{eq:defTauAllRelations}, \ref{eq:defTauNab}).  As discussed above, there are 502 joint constraints for quantities $\tau_{lr}$ in each basis for this protocol.  In principle, one can solve this optimization problem by the linear programming (LP) method with all those 502 constraints being listed in Eqs.(\ref{eq:defTauAllRelations}, \ref{eq:defTauNab}). However, this will cost huge computation power in making full optimization of all parameters. Note that if we only use part of the constraints, the final key will be still secure but the key rate could be not the optimized result. Naturally, one may ask the question whether we can still obtain the optimized or almost optimized result if we only use a few of 502 constraints. The answer is yes.  Actually, as shown below, most of the constrains take no effect to the key rate and they can be abandoned. Luckily, as we show below, we can greatly reduce the number of joint constraints to 11 or even fewer. Moreover, we can even obtain explicit formulas for the optimization problem.

Reconsidering the function $e_{11}^{f}$, we know that the signs in front of $T_{+}^{f}$ and $T_{-}^{f}$ are different. So we can treat the variables in $T_{+}^{f}$ and $T_{-}^{f}$ separately. That is to say, equivalently, the maximization of $e_{11}^{f}$ can be divided into two simple problems that are the maximization of $T_{+}^{f}$ and the minimization of $T_{-}^{f}$. In maximizing $T_{+}^{f}$, we only need to consider the sole joint constraint between variables $\tau_{xx}$ and $\tau_{oo}$. Similarly, in minimizing $T_{-}^{f}$, we only need to consider the sole joint constraint between variables $\tau_{ox}$ and $\tau_{xo}$. These optimization problems can be solved by using the LP method. Furthermore, we can solve the problem analytically with explicit formulas.
We consider the maximization of $T_{+}^{f}$ first. In this function, there are two variables $\tau_{xx}$ and $\tau_{oo}$. So we only need to consider the joint constraint $N_{xx} T_{xx} \tau_{xx} +N_{oo} T_{oo}\tau_{oo} \leq n_{\tau}\sqrt{N_{xx} T_{xx}+N_{oo} T_{oo}}$ and the bounds of $\tau_{xx}$, $\tau_{oo}$ shown in Eq.(\ref{eq:defTau}). Explicitly, we have
\begin{eqnarray}\label{eq:TpXf1}
  T_{+}^{f}&\leq& \hat{T}_{+}+\frac{n_{\tau} a_0 b_0}{N_{oo}}\sqrt{ N_{xx}T_{xx} + N_{oo}T_{oo}} \nonumber \\
  & & \hspace{-5mm} +n_{\tau} \left(\frac{1}{N_{xx}}-\frac{a_0 b_0}{N_{oo}}\right)\sqrt{N_{xx} T_{xx}} \triangleq \overline{T}_{+}^{f_1},
\end{eqnarray}
when $a_0 b_0 N_{xx}\leq N_{oo}$, and
\begin{eqnarray}\label{eq:TpXf2}
  T_{+}^{f}&\leq& \hat{T}_{+}+\frac{n_{\tau} }{N_{xx}}\sqrt{ N_{xx}T_{xx} + N_{oo}T_{oo}} \nonumber \\
  & & \hspace{-5mm} +n_{\tau} \left(\frac{a_0 b_0}{N_{oo}}-\frac{1}{N_{xx}}\right)\sqrt{N_{oo} T_{oo}} \triangleq \overline{T}_{+}^{f_2},
\end{eqnarray}
otherwise.

Conclusively, according to Eq.(\ref{eq:TpXf1}) and Eq.(\ref{eq:TpXf2}), the upper bound of the function $T_{+}^{f}$ can be defined as $\overline{T}_{+}^{f}=\overline{T}_{+}^{f_1}$ when $a_0 b_0 N_{xx}\leq N_{oo}$ and $\overline{T}_{+}^{f}=\overline{T}_{+}^{f_2}$ otherwise. Furthermore, we can easily prove that the upper bound $\overline{T}_{+}^{f}$ is reachable. So $\overline{T}_{+}^{f}$ is just the maximum value of the function ${T}_{+}^{f}$ under the joint constraints.

For convenience, we can write $\overline{T}_{+}^{f}$ uniformly by introducing the following notations
\begin{equation*}
  d_{xx}=1/N_{xx}, \quad d_{oo}=a_0 b_0/N_{oo}.
\end{equation*}
Moreover, we use nature numbers $1,2$ to indicate these two different subscripts $xx$ and $oo$ with ascending order of $d_{k}(k=1,2)$. That is to say, we use nature number $1,2$ to indicate $xx$, $oo$ respectively when $d_{xx}\leq d_{oo}$, and to indicate $oo$, $xx$ respectively when $d_{oo}\leq d_{xx}$. With these preparations, we can write the maximum value of $T_{+}^{f}$ uniformly
\begin{eqnarray}\label{eq:e11stapAna}
  \overline{T}_{+}^{f}&=& \hat{T}_{+}+n_{\tau} d_1\sqrt{ N_{1}T_{1} + N_{2}T_{2}} \nonumber \\
  & & +n_{\tau}(d_2-d_1)\sqrt{N_{2} T_{2}}.
\end{eqnarray}

Similarly, we can minimize the function $T_{-}^{f}$ with the boundary constraints $\tau_{ox}\geq -\overline{\tau}_{ox}$, $\tau_{xo}\geq -\overline{\tau}_{xo}$ and the joint constraint $N_{ox} T_{ox} \tau_{ox} +N_{xo} T_{xo}\tau_{xo} \geq -n_{\tau}\sqrt{N_{ox} T_{ox}+N_{xo} T_{xo}}$. In the symmetric case where $a_k=b_k$, $a_k^{\prime}=b_k^{\prime}$, and $N_{lr}=N_{rl}$, the minimum value of $T_{-}^{f}$ can be easily obtained
\begin{equation}\label{eq:TmXfL1}
  T_{-}^{f}\geq \hat{T}_{-}-\frac{n_{\tau} a_0}{N_{ox}}\sqrt{N_{ox} T_{ox} +N_{xo} T_{xo}} \triangleq \underline{T}_{-}^{f},
\end{equation}
where we have used the symmetric conditions and the joint constraint between $\tau_{ox}$ and $\tau_{xo}$. Generally, without the symmetric case assumption, we can write the minimum value of $T_{-}^{f}$ into
\begin{eqnarray}\label{eq:e11stanAna}
  \underline{T}_{-}^{f}&=& \hat{T}_{-}-n_{\tau} d_1\sqrt{ N_{1}T_{1} + N_{2}T_{2}} \nonumber \\
  & & -n_{\tau}(d_2-d_1)\sqrt{N_{2} T_{2}},
\end{eqnarray}
where we use nature numbers $1,2$ to indicate these two different subscripts $ox$, $xo$ with ascending order of $d_k(k=1,2)$ and $d_{ox}=a_0/N_{ox}$, $d_{xo}=b_0/N_{xo}$.

With the maximum value $\overline{T}_{+}^{f}$ and the minimum value $\underline{T}_{-}^{f}$, we can define the upper bound of $e_{11}$ with the following explicit formula
\begin{equation}\label{eq:e11staAna}
  \overline{e}_{11}^{f}=(\overline{T}_{+}^{f}-\underline{T}_{-}^{f})/(a_1 b_1 \underline{s}_{11}^{f}),
\end{equation}
where $\overline{T}_{+}^{f}$ and $\underline{T}_{-}^{f}$ are defined in Eq.(\ref{eq:e11stapAna}) and Eq.(\ref{eq:e11stanAna}) respectively, $\underline{s}_{11}^{f}$ is the lower bound of $s_{11}$ that will be studied in the coming.

Now we commit ourself to derive the explicit formula to estimate the lower bound of $s_{11}$. As is defined in Eq.(\ref{eq:ExptS}), the mean values $\langle{S}_{lr}\rangle$ is the function of  $\delta_{lr}$. Replacing $S_{lr}$ by its mean value $\langle{S}_{lr}\rangle$ in Eq.(\ref{eq:s11of14}), we get a function about quantities $\delta_{lr}$
\begin{equation}\label{eq:s11of14fab}
  s_{11}^{f_1}=\hat{\underline{s}}_{11}^{(1)}+(S_{+}^{f_1}-S_{-}^{f_1})/(a_1 a_1^{\prime} \tilde{b}_{12}),
\end{equation}
where $S_{+}^{f_1}=\mathcal{S}_{+}^{(1)}(S_{xx}\delta_{xx}, S_{oy}\delta_{oy}, S_{yo}\delta_{yo}, S_{oo}\delta_{oo})$, $S_{-}^{f_1}=\mathcal{S}_{-}^{(1)}(S_{yy}\delta_{yy}, S_{ox}\delta_{ox}, S_{xo}\delta_{xo})$  with $\mathcal{S}_{+}^{(1)}$, $\mathcal{S}_{-}^{(1)}$ being defined in Eqs.(\ref{eq:DefSp1},\ref{eq:DefSn1}) respectively,
and $\hat{\underline{s}}_{11}^{(1)}$ being the constant factor defined in Eq.(\ref{eq:s11of14}).

To obtain the lower bound of $s_{11}$ from this function $s_{11}^{f_1}$, we need to find out the worst case under the constraints about the quantities $\delta_{lr}$. That is to say, we need minimize the function $s_{11}^{f_1}$ of variables $\delta_{lr}$ under the constraints shown in Eqs.(\ref{eq:defDeltaAllRelations},\ref{eq:defDeltaNab}). In the function $s_{11}^{f_1}$, we can see that the signs in front of $S_{+}^{f_1}$ and $S_{-}^{f_1}$ are different. So we can treat the variables in $S_{+}^{f_1}$ and $S_{-}^{f_1}$ separately. That is to say, equivalently, the minimization of $s_{11}^{f_{1}}$ can be divided into two simple problems that are the minimization of $S_{+}^{f_1}$ and the maximization of $S_{-}^{f_1}$. In minimizing $S_{+}^{f_1}$, we only need to consider the constraints among variables $\delta_{xx}$, $\delta_{oy}$, $\delta_{yo}$ and $\delta_{oo}$. There are only 11 joint constraints in this LP problem. Similarly, in maximizing $S_{-}^{f_1}$, we only need to consider the constraints among variables $\delta_{yy}$, $\delta_{ox}$ and $\delta_{xo}$. Here there are only 4 joint constraints.

Similar to the upper bound of $e_{11}$,  we can also lower bound $s_{11}$ analytically.  For clarity, we introduce the following theorem
\begin{theorem}\label{thm:LPSolver}
  Consider the $K-$variable linear function $f(x_k)=\sum_{k=1}^{K} \alpha_k x_k$ with $x_k(k=1,2,\cdots,K;K\leq 4)$ and the following linear constraints
  \begin{equation}\label{eq:LPConstraints}
    \left|\sum_{k\in\mathcal{K}}\beta_k x_k\right|\leq n_0\sqrt{\sum_{k\in \mathcal{K}} \beta_k}, \quad \mathcal{K}\subseteq \{1,2,\cdots,K\}
  \end{equation}
  where $n_0\geq 0$ and $\alpha_k,\beta_k\geq 0$ $(k=1,2,\cdots,K)$. The maximum value of $f(x_k)$ is
  \begin{eqnarray}\label{eq:LPfmax}
    f_{max}&=&f(\tilde{x}_{k}^{*})=\mathcal{F}(K,n_0,V_{\alpha},V_{\beta}) \nonumber \\
    &=&n_0 \sum_{k=1}^{K}(\tilde{\gamma}_n -\tilde{\gamma}_{n-1})\sqrt{\sum_{k=n}^{K}\tilde{\beta}_n},
  \end{eqnarray}
  with
  \begin{equation}\label{eq:LPxk}
    \tilde{x}_{k}^{*}=\frac{n_0}{\tilde{\beta}_k}\left(\sqrt{\sum_{n=k}^{K} \tilde{\beta}_n} -\sqrt{\sum_{n=k+1}^{K}\tilde{\beta}_n}\right), \, (k=1,\cdots,K)
  \end{equation}
  and the minimum value of $f(x_k)$ is
  \begin{equation}\label{eq:LPfmin}
    f_{min}=-f_{max}=f(-\tilde{x}_k^{*})=-\mathcal{F}(K,n_0,V_{\alpha},V_{\beta})
  \end{equation}
  given the following notations:
   1, $V_{\alpha}=[\alpha_1,\alpha_2,\cdots,\alpha_K]$, $V_{\beta}=[\beta_1,\beta_2,\cdots,\beta_K]$, $\tilde{\gamma}_0=0$, $\gamma_k=\alpha_k/\beta_k$ and $\tilde{\gamma}_k$ is the $k$-th minimum value of $\{\gamma_k|k=1,2,\cdots,K\}$ which means that we use $\tilde{\gamma}_k$ to denote the values of $\{\gamma_k\}$ with ascending order. 2, $\tilde{\zeta}_k$ is the rearrangement of $\zeta_k (\zeta=\alpha,\beta,x)$ so that $\tilde{\gamma}_k=\tilde{\alpha}_k/\tilde{\beta}_k(k=1,2,\cdots,K)$.
\end{theorem}

In Eq.(\ref{eq:LPConstraints}), we use $\mathcal{K}$ to denote the subsects of $\{1,2,\cdots,K\}$. Then we know that there are $\sum_{k=2}^{K}\mathcal{C}_{K}^{k}$ joint constraints and $\mathcal{C}_K^1=K$ boundary constraints about $x_k (k=1,2,\cdots,K)$. For example, in the case with $K=4$, we have $11$ joint constraints such as $|\beta_1 x_1 +\beta_2 x_2 +\beta_4 x_4|\leq n_0\sqrt{\beta_1+\beta_2 +\beta_4}$ with $\mathcal{K}=\{1,2,4\}$ and $4$ boundary constraints such as $|x_1|\leq n_0/\sqrt{\beta_1}$ with $\mathcal{K}=\{1\}$. Actually, the value $f_{max}$ is an upper bound of the function $f(x_k)$. On the other hand, we can prove that the point $P_s=(\tilde{x}_1^{*}, \tilde{x}_2^{*}, \tilde{x}_3^{*}, \tilde{x}_4^{*})$ locates in the feasible region. Details of the proof of this theorem can be see in Appendix~\ref{app:ThmProof}.

By using the conclusion shown in Theorem~\ref{thm:LPSolver} with $K=2$, we can easily find out the maximum value of $T_{+}^{f}$ and the minimum value of $T_{-}^{f}$. These values have been shown above in Eq.(\ref{eq:e11stapAna}) and Eq.(\ref{eq:e11stanAna}) respectively.

As discussed above, the function $S_{+}^{f_1}$ contains four variables $\delta_{xx}$, $\delta_{oy}$, $\delta_{yo}$ and $\delta_{oo}$. In order to find out the minimum value of $S_{+}^{f_1}$ with the constraints shown in Eq.(\ref{eq:defDeltaAllRelations}), we can use Theorem~\ref{thm:LPSolver} with $K=4$ directly. Explicitly, we have
\begin{equation}\label{eq:SpfaLv}
  S_{+}^{f_1}\geq - \mathcal{F}(4,n_{\delta},V_{\alpha}^{+},V_{\beta}^{+}) \triangleq \underline{S}_{+}^{f_1},
\end{equation}
where the function $\mathcal{F}$ is defined in Eq.(\ref{eq:LPfmax}) and $V_{\alpha}^{+}=[a_1^{\prime}b_2^{\prime} S_{xx}, \allowbreak a_1 b_2 a_0^{\prime} S_{oy}, \allowbreak a_1 b_2 b_0^{\prime} S_{yo},\allowbreak  (a_1^{\prime} b_2^{\prime} a_0 b_0-a_1 b_2 a_0^{\prime} b_0^{\prime})S_{oo}]$, $V_{\beta}^{+}=[N_{xx}S_{xx},\allowbreak  N_{oy} S_{oy}, \allowbreak N_{yo}S_{yo}, \allowbreak N_{oo}S_{oo}]$.

Similarly, for the function $S_{-}^{f_1}$ with three variables $\delta_{yy}$, $\delta_{ox}$ and $\delta_{xo}$, we can estimate the maximum value of it by using Theorem~\ref{thm:LPSolver} with $K=3$
\begin{equation}\label{eq:SnfaUv}
  S_{-}^{f_1}\leq \mathcal{F}(3,n_{\delta},V_{\alpha}^{-},V_{\beta}^{-})  \triangleq \overline{S}_{-}^{f_1},
\end{equation}
where the function $\mathcal{F}$ is defined in Eq.(\ref{eq:LPfmax}) and $V_{\alpha}^{-}=[a_1 b_2 S_{yy}, \allowbreak a_1^{\prime} b_2^{\prime} a_0 S_{ox}, \allowbreak a_1^{\prime} b_2^{\prime} b_0 S_{xo}]$, $V_{\beta}^{-}=[N_{yy}S_{yy}, \allowbreak N_{ox} S_{ox},\allowbreak  N_{xo}S_{xo}]$.

After these preparations, we obtain a lower bound of $s_{11}$ with the explicit formula
\begin{equation}\label{eq:s11of14staAna}
  \underline{s}_{11}^{f_1}=\hat{s}_{11}^{(1)}+(\underline{S}_{+}^{f_1}-\overline{S}_{-}^{f_1})/(a_1 a_1^{\prime} \tilde{b}_{12}),
\end{equation}
where $\hat{s}_{11}^{(1)}$ is defined in Eq.(\ref{eq:s11of14}), $\underline{S}_{+}^{f_1}$ is the minimum values of $S_{+}^{f_1}$, $\overline{S}_{-}^{f_1}$ is the maximum values of $S_{-}^{f_1}$ which are shown in Eqs.(\ref{eq:SpfaLv},\ref{eq:SnfaUv}) respectively.

Besides the lower bound $\underline{s}_{11}^{f_1}$, we can obtain another lower bound of $s_{11}$ corresponding to $\underline{\hat{s}}_{11}^{(2)}$ shown in Eq.(\ref{eq:s11of123}). Replacing $S_{lr}$ by its mean value $\langle{S}_{lr}\rangle$ in Eq.(\ref{eq:s11of123}), we get a function about quantities $\delta_{lr}$
\begin{equation}\label{eq:s11of123stafun}
  {s}_{11}^{f_2}=\underline{\hat{s}}_{11}^{(2)}+\frac{S_{+}^{f_2}-S_{-}^{f_2}}{a_1 b_1 \tilde{a}_{12}\tilde{b}_{12}},
\end{equation}
where $\underline{\hat{s}}_{11}^{(2)}$ is a constant factor given by Eq.(\ref{eq:s11of123}), $\tilde{a}_{12}=a_1 a_2^{\prime} -a_1^{\prime}a_2$, $\tilde{b}_{12}=b_1 b_2^{\prime} -b_1^{\prime}b_2$ and $S_{+}^{f_2}=\mathcal{S}_{+}^{(2)}(S_{xx}\delta_{xx}, S_{oy}\delta_{oy}, S_{yo}\delta_{yo}, S_{oo}\delta_{oo})$, $S_{-}^{f_2}=\mathcal{S}_{-}^{(2)}(S_{xy}\delta_{xy}, S_{yx}\delta_{yx}, S_{ox}\delta_{ox}, S_{xo}\delta_{xo})$ with $\mathcal{S}_{+}^{(2)}$, $\mathcal{S}_{-}^{(2)}$ being defined in Eqs.(\ref{eq:s11of123Sp},\ref{eq:s11of123Sn}) respectively.

Similarly to the lower bound of $\underline{s}_{11}^{f_1}$, we need to find out the minimum value of $S_{+}^{f_2}$ and the maximum value of $S_{-}^{f_2}$ with the constraints shown in Eq.(\ref{eq:defDeltaAllRelations}).

Using theorem 1, we can also find out the explicit formulas for the minimum of $S_{+}^{f_2}$ and maximum of $S_{-}^{f_2}$ analytically
\begin{eqnarray}
  S_{+}^{f_2}&\geq& - \mathcal{F}(4,n_{\delta},W_{\alpha}^{+},W_{\beta}^{+}) \triangleq \underline{S}_{+}^{f_2}, \label{eq:Spf2Lv}  \\
  S_{-}^{f_2}&\leq& \mathcal{F}(4,n_{\delta},W_{\alpha}^{-},W_{\beta}^{-}) \triangleq \overline{S}_{-}^{f_2}, \label{eq:Snf2Uv}
\end{eqnarray}
where the function $\mathcal{F}$ is defined in Eq.(\ref{eq:LPfmax}) and $W_{\alpha}^{+}=[g_{xx} S_{xx}, \allowbreak g_{oy} S_{oy}, \allowbreak g_{yo} S_{yo}, \allowbreak g_{oo} S_{oo}]$, $W_{\beta}^{+}=V_{\beta}^{+}$, $W_{\alpha}^{-}=[g_{xy} S_{xy}, \allowbreak  g_{yx} S_{yx}, \allowbreak g_{ox} S_{ox}, \allowbreak g_{xo} S_{xo}]$, $W_{\beta}^{-}=[N_{xy}S_{xy}, \allowbreak N_{yx} S_{yx}, \allowbreak N_{ox}S_{ox}, \allowbreak N_{xo} S_{xo}]$.

With these estimations, we can write the other lower bound of $s_{11}$ into the explicit formula
\begin{equation}\label{eq:s11of123staAna}
  \underline{s}_{11}^{f_2}=\underline{\hat{s}}_{11}^{(2)} + \frac{\underline{S}_{+}^{f_2}-\overline{S}_{-}^{f_2}}{a_1 b_1 \tilde{a}_{12}\tilde{b}_{12}},
\end{equation}
where $\underline{\hat{s}}_{11}^{(2)}$ is a constant factor defined by Eq.(\ref{eq:s11of123}), $\underline{S}_{+}^{f_2}$ is the minimum value of $S_{+}^{f_2}$ and $\overline{S}_{-}^{f_2}$ is the maximum value of $S_{-}^{f_2}$ which are defined in Eqs.(\ref{eq:Spf2Lv},\ref{eq:Snf2Uv}) respectively.

Combining these two lower bounds, we obtain our improved estimation of the lower bound of $s_{11}$. Explicitly, we define
\begin{equation}\label{eq:s11staAna}
  \underline{s}_{11}^{f}=\max\{ \underline{s}_{11}^{f_1}, \underline{s}_{11}^{f_2}\},
\end{equation}
where $\underline{s}_{11}^{f_1}$ and $\underline{s}_{11}^{f_2}$ are defined in Eq.(\ref{eq:s11of14staAna}) and Eq.(\ref{eq:s11of123staAna}) respectively.

So far we have completed our analytical formulas. As discussed above, result of our analytical method is the same with that of the LP method which fully uses all those $502$ joint constraints.

\section{Numerical simulation}\label{sec:SimulationNew}
In this section, we will present some results of numerical simulations. We treat the statistical fluctuation jointly as studied earlier in this paper. We also optimize all parameters by the method of full optimization\cite{Xu1406}. We shall compare our results with Ref.\cite{Xu1406}, which has proposed the method of full optimization and has presented obviously the best result among all prior art works. We shall also compare our results with the result obtained by what we called {\em Traditional} method, i.e., optimizing all parameters but treating the statistical fluctuations of different sources {\em separately}. Without any loss of generality, we focus on the symmetric case where the two channel transmissions from Alice to UTP and from Bob to UTP are equal. Therefore we set $p_{l}=p_r$  and also $p_{X|l}=p_{X|r}$ for any $l=r$. We also assume that the UTP's detectors are identical, i.e., they have the same dark count rates and detection efficiencies, and their detection efficiencies do not depend on the incoming signals. We shall estimate what values would be probably observed for the yields and error yields in the normal cases by the linear channel loss model as in~\cite{wang05,MaPRA2012,ind2,WangModel}. For a fair comparison, we use the same experimental parameters used in Ref.\cite{Xu1406} for our numerical simulation, which are mostly from the long-distance QKD experiment reported in~\cite{UrsinNP2007}. The values of these parameters are listed in Table~\ref{tab:parameters}. With this, the yields $S_{lr}$ and error yields $T_{lr}$ can be calculated~\cite{WangModel} with coherent states with intensities $\mu_{o},\mu_x,\mu_y$. The density matrix of the coherent state with intensity $\mu$ can be written into $\rho=\sum_{k}\frac{e^{-\mu}\mu^k}{k!} \oprod{k}{k}$. By using these values, we can estimate the lower bounds of $s_{11}^{Z}$ and the upper bounds of $e_{11}^{X}$ with different methods presented in the above sections. With these preparations, we can calculate the final secret key rate with the following formula~\cite{ind2}
\begin{equation}\label{eq:KeyRate}
  R=\frac{1}{2}p_{y}^2p_{Z|y}^2a_1^{\prime} b_1^{\prime} s_{11}^{Z}[1-H(e_{11}^{X})]-f_e S_{y y}^{Z}H(E_{y y}^{Z}),
\end{equation}
where $S_{y y}^{Z}$ and $E_{y y}^{Z}$ denote, respectively, the yield and error rate in $Z$-basis when Alice uses ${y_A}$ and Bob uses ${y_B}$; $f_e$ is the efficiency factor of the error correction method used; $s_{11}^{Z}$ and $e_{11}^{X}$ are the yield and error rate when both Alice and Bob send single-photon states.

\begin{table}
\begin{ruledtabular}
\begin{tabular}{cccccc}
  $e_0$ & $e_d$ & $p_d$ & $\eta_d$ & $f_e$ & $\epsilon$  \\
  \hline
  0.5 & 1.5\% & $6.02\times 10^{-6}$ & $14.5\%$ & $1.16$ & $1.0\times 10^{-7}$ \\
\end{tabular}
\caption{\label{tab:parameters}List of experimental parameters used in numerical simulations. $e_0$ is the error rate of background. $e_d$ is the misalignment-error probability. $p_d$ is the dark count rate. $\eta_d$ is the detection efficiency of all detectors. $f_e$ is the error correction inefficiency. $\epsilon$ is the security bound considered in the finite-date analysis.}
\end{ruledtabular}
\end{table}

To make a fair comparison, we need to find out the full parameter optimizations for different methods~\cite{Xu1406}. Here we also use the well-known local search algorithm. In this algorithm, we need to optimize the one-variable nonlinear function in each step for the local search.

We consider the three-intensity protocol in the case of data-size $N_t=10^{12}$. The optimal parameters and the practical key rate per pulse for the distance 50km (standard fiber), with the statistical fluctuations, are shown in Table~\ref{tab:3Int50kmNt12}. The result presented in Ref.\cite{Xu1406} is shown in the 2nd column. In the 3rd column, we show the optimal parameters after a full parameter optimization by using the traditional analytical method with Eq.(\ref{eq:s11sta}) and Eq.(\ref{eq:e11sta}). In the 4th column, we present the results after a full parameter optimization by using the improved analytical method with explicit formulas in Eq.(\ref{eq:s11staAna}) and Eq.(\ref{eq:e11staAna}). We can see that our new optimal key rate with full parameter optimization is better than the result presented in Ref.\cite{Xu1406}. The improvement of the optimal final key rate $R$ is about $97\%$. In Ref.\cite{Xu1406}, the lowest intensity is $10^{-6}$ which is too small to be obtained exactly in real experimental. On the other hand, the optimal key rates obtained by setting the lowest intensity to be 0 or $10^{-6}$ respectively are nearly equal to each other. Furthermore, the improved method considered in this paper can apply to the generalized situation that the lowest intensity is not 0 \cite{Wang3g}. In Table~\ref{tab:3Int100kmNt12}, we show the optimal parameters and the practical key rate per pulse for the distance 100km (standard fiber) with the statistical fluctuations. In the 2nd, 3rd and 4th columns, we present the results after a full parameter optimization by using the traditional method, the improved method with formulas and the LP method with all joint constraints for statistical fluctuations. We can see that our new full parameter optimization can improve the key rate $R$ by $146\%$.

\begin{table}
\begin{ruledtabular}
\begin{tabular}{c|ccc}
  \rm{Parameter} & Ref.\cite{Xu1406} & \rm{Traditional} & \rm{Improved}  \\
  \hline
  $\mu_y$ & 0.25 & 0.396 & 0.401 \\
  $\mu_x$ & 0.05 & 0.056 & 0.055 \\
  $\mu_o$ & $1.0\times 10^{-6}$ & $0$ & $0$  \\
  \hline
  $p_y$   & 0.58 & 0.646 & 0.681 \\
  $p_x$   & 0.30 & 0.256 & 0.243 \\
  \hline
  $p_{X|y}$ & 0.03 & 0.024 & 0.013 \\
  $p_{X|x}$ & 0.71 & 0.737 & 0.709 \\
  $p_{X|o}$ & 0.83 & 1.000 & 1.000 \\
  \hline
  $R$     & $1.68\times 10^{-6}$ & $2.59\times 10^{-6}$ & $3.31\times 10^{-6}$
\end{tabular}
\caption{\label{tab:3Int50kmNt12} Comparison of parameters at 50kms (standard fiber) for three-intensity protocol with statistical fluctuation analysis in the case of data-size (total number of pulse pairs) $N_t=10^{12}$. The 2nd column is the result presented in Ref.\cite{Xu1406}. The 3rd column is the optimal parameters after a full parameter optimization with explicit formulas Eq.(\ref{eq:s11sta}) and Eq.(\ref{eq:e11sta}). The 4th column is the optimal parameters obtained with explicit formulas Eq.(\ref{eq:s11staAna}) and Eq.(\ref{eq:e11staAna}). Comparing with the result given by Xu et al\cite{Xu1406}, our improved result raises the key rate $R$ by $97\%$. We set the same probabilities for Alice and Bob in choosing sources and bases, $p_x, p_y$: probabilities to choose source $x$ and $y$; $p_{X|y}$, $p_{X|x}$,
$p_{X|o}$ : probabilities of choosing $X-$basis selection conditional on sources $y$, $x$ and $o$. Note that in Ref.\cite{Xu1406}, source $o$ is not strict vacuum.}
\end{ruledtabular}
\end{table}

\begin{table}
\begin{ruledtabular}
\begin{tabular}{c|ccc}
  \rm{Parameter} & \rm{Traditional} & \rm{Improved I} & Improved II \\
  \hline
  $\mu_y$ & 0.269 & 0.275 & 0.275 \\
  $\mu_x$ & 0.067 & 0.068 & 0.068 \\
  \hline
  $p_y$   & 0.336 & 0.404 & 0.404 \\
  $p_x$   & 0.477 & 0.447 & 0.447 \\
  \hline
  $p_{X|y}$ & 0.132 & 0.084 & 0.084 \\
  $p_{X|x}$ & 0.742 & 0.719 & 0.720\\
  \hline
  $R$     & $1.00\times 10^{-8}$ & $2.46\times 10^{-8}$ & $2.46\times 10^{-8}$
\end{tabular}
\caption{\label{tab:3Int100kmNt12} Comparison of parameters at 100kms (standard fiber) for three-intensity protocol with statistical fluctuation analysis with $N_t=10^{12}$. The 2nd column is the optimal parameters after a full parameter optimization with explicit formulas Eq.(\ref{eq:s11sta}) and Eq.(\ref{eq:e11sta}). The 3rd column is the optimal parameters obtained by our improved method with explicit formulas Eq.(\ref{eq:s11staAna}) and Eq.(\ref{eq:e11staAna}). The 4th column is the optimal parameters obtained by the LP method with all 502 joint constraints. We can see that our new full parameter optimization can improve the key rate $R$ by $146\%$. The results obtained by using the numerical method and analytical methods are near equal to each other.}
\end{ruledtabular}
\end{table}

More extensive comparison results are shown in Fig.1. In this figure, we show the optimal key rate (per pulse) in logarithmic scale as a function of the distance under a practical setting with finite data-set $N_t=10^{12}$.
\begin{figure}
  \includegraphics[width=240pt]{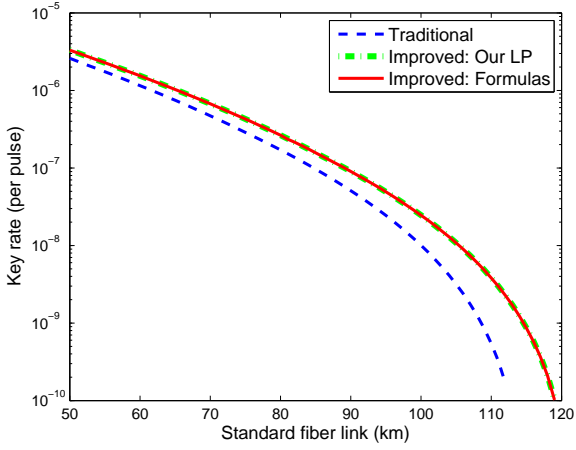}\\
  \caption{(Color online) Optimal secret key rate (per pulse) as a function of the distance under different methods of three-intensity decoy-state protocol. Here we set $N_t=10^{12}$. Dashed curve: key rates obtained through traditional full optimization with simple worst-case treatment for fluctuations in Eq.(\ref{eq:s11sta}) and Eq.(\ref{eq:e11sta}); Dashed-dotted curve: results from the LP method which fully used all 502 joint constraints of fluctuations;  Solid curve: key rates from our formulas Eq.(\ref{eq:s11staAna}) and Eq.(\ref{eq:e11staAna}) with joint treatment of fluctuations.  In the simulation, we have taken a full parameter optimization for all points.}\label{ROptNt12}
\end{figure}

\section{Conclusion}\label{sec:Conclusion}
Through studying the statistical fluctuations of different sources jointly, we obtain the improved statistical analysis with explicit formulas. Numerical simulation shows that the results obtained from our improved methods are significantly better than the results obtained with the traditional methods treating statistical fluctuations of each sources separately. In our study, we have taken the same intensities in both bases. The result can be further improved by taking different intensities in different bases and taking the decoy-state method in only one basis as pointed out in \cite{Wang3improve}. This will be reported elsewhere.
\\{\bf Acknowledgement} We
acknowledge the financial support in part by the 10000-Plan of Shandong province, and the
National High-Tech Program of China grant No. 2011AA010800 and 2011AA010803,
NSFC grant No. 11474182, 11174177 and 60725416.

\appendix
\section{The proof of Theorem~\ref{thm:LPSolver}}\label{app:ThmProof}
In section~\ref{sec:Improved}, we need to use the conclusion of Theorem~\ref{thm:LPSolver} with $K=2,3,4$. Actually, we only need to prove the result with $K=4$ since the cases with $K=2,3$ can be treated as the special case with $K=4$. Furthermore, we only need to prove the situation in maximizing $f(x_k)$.

First, we show that the value $f_{max}$ defined in Eq.(\ref{eq:LPfmax}) is an upper bound of the function $f(x_k)$. Actually, the function $f(x_k)$ can be written into
\begin{eqnarray*}
  f(x_k)&=& \sum_{k=1}^{4} \gamma_k \beta_k x_k =\sum_{k=1}^{4} \tilde{\gamma}_k \tilde{\beta}_k \tilde{x}_k \nonumber \\
  &=&\sum_{n=1}^{4}(\tilde{\gamma}_n-\tilde{\gamma}_{n-1}) \sum_{k=1}^{4} \tilde{\beta}_k \tilde{x}_k,
\end{eqnarray*}
where $\gamma_k$, $\tilde{\gamma}_k$, $\tilde{\alpha}_k$, $\tilde{\beta}_k$ and $\tilde{x}_k$ are defined in Theorem~\ref{thm:LPSolver}. Accordingly, we know that $\tilde{\gamma}_1\leq \tilde{\gamma}_2\leq \tilde{\gamma}_3\leq \tilde{\gamma}_4$. So we have $\tilde{\gamma}_n-\tilde{\gamma}_{n-1}\geq 0$ $(n=1,2,3,4)$. On the other hand, we know that $\sum_{k=n}^{4}\tilde{\beta}_k \tilde{x}_k \leq n_0\sqrt{\sum_{k=n}^{4}\tilde{\beta}_k}$ with the constraints shown in Eq.(\ref{eq:LPConstraints}). Then we can conclude that
\begin{equation*}
  f(x_k)\leq n_0 \sum_{n=1}^{4}(\tilde{\gamma}_n-\tilde{\gamma}_{n-1}) \sqrt{\sum_{k=1}^{4} \tilde{\beta}_k }=f_{max}.
\end{equation*}

Secondly, we prove that the upper bound $f_{max}$ is reachable. As discussed above, we have used four constraints $\sum_{k=n}^{4}\tilde{\beta}_k \tilde{x}_k \leq n_0 \sqrt{\sum_{k=n}^{4}\tilde{\beta}_k}$ $(n=1,2,3,4)$ in obtaining the upper bound $f_{max}$. Then we can solve the linear systems $\sum_{k=n}^{4}\tilde{\beta}_k \tilde{x}_k = n_0 \sqrt{\sum_{k=n}^{4}\tilde{\beta}_k}$ $(n=1,2,3,4)$ about variables $\tilde{x}_k$ with
\begin{equation*}
  \tilde{x}_{k}^{*}=\frac{n_0}{\tilde{\beta}_k}\left(\sqrt{\sum_{n=k}^{4} \tilde{\beta}_n} -\sqrt{\sum_{n=k+1}^{4}\tilde{\beta}_n}\right), \, (k=1,2,3,4).
\end{equation*}
In order to prove the upper bound $f_{max}=f(\tilde{x}_k^{*})$ is reachable, we only need to show that the point $P_s=(\tilde{x}_1^{*}, \tilde{x}_2^{*}, \tilde{x}_3^{*}, \tilde{x}_4^{*})$ locates in the feasible region. That is to say, we need to prove that all the constraints presented in Eq.(\ref{eq:LPConstraints}) with $K=4$ are fulfilled when $\tilde{x}_k=\tilde{x}_k^{*}$. Equivalently, the constraints in Eq.(\ref{eq:LPConstraints}) with $K=4$ can be rewritten into
\begin{equation}\label{eq:LPConstraintsK4}
  \sum_{k\in\mathcal{K}}\tilde{\beta}_k \tilde{x}_k \leq n_0\sqrt{\sum_{k\in \mathcal{K}} \tilde{\beta}_k}, \quad \mathcal{K}\subseteq \{1,2,3,4\}.
\end{equation}

There are $\sum_{k=1}^{4}\mathcal{C}_4^k=15$ constraints in Eq.(\ref{eq:LPConstraintsK4}). In the following, we will group these 15 cases into 4 situations. In the first situation, we consider the constraints with $\mathcal{K}=\mathcal{K}_{n}^{(1)} (n=1,2,3,4)$ and $\mathcal{K}_1^{(1)}=\{4\}$, $\mathcal{K}_2^{(1)}=\{3,4\}$, $\mathcal{K}_3^{(1)}=\{2,3,4\}$, $\mathcal{K}_4^{(1)}=\{1,2,3,4\}$. In this situation, we can easily calculate that
\begin{equation*}
  \sum_{k\in\mathcal{K}_n^{(1)}}\tilde{\beta}_k \tilde{x}_k = n_0\sqrt{\sum_{k\in \mathcal{K}_n^{(1)}} \tilde{\beta}_k}, \quad (n=1,2,3,4).
\end{equation*}

In the second situation, we consider the constraints with $\mathcal{K}=\mathcal{K}_{n}^{(2)} (n=1,2,\cdots,6)$ and $\mathcal{K}_1^{(2)}=\{1\}$, $\mathcal{K}_2^{(2)}=\{2\}$, $\mathcal{K}_3^{(2)}=\{3\}$, $\mathcal{K}_4^{(2)}=\{1,2\}$, $\mathcal{K}_5^{(2)}=\{2,3\}$, $\mathcal{K}_6^{(2)}=\{1,2,3\}$. In this situation, the following mean value theorem should be used
\begin{equation*}
  \sqrt{y_1+y_2}\leq \sqrt{y_1}+\sqrt{y_2}, \quad (y_1,y_2\geq 0).
\end{equation*}
Taking $\mathcal{K}=\mathcal{K}_{5}^{(2)}$ as an example, we have
\begin{equation*}
  \sum_{k\in \mathcal{K}_5^{(2)}}\tilde{\beta}_k \tilde{x}_k =n_0 \sqrt{\tilde{\beta}_2 + \tilde{\beta}_3 + \tilde{\beta}_4} -n_0 \sqrt{\tilde{\beta}_4} \leq n_0 \sqrt{\tilde{\beta}_2 + \tilde{\beta}_3}.
\end{equation*}
We can also prove the other cases in the same way.

In the third situation, we consider the constraints with $\mathcal{K}=\mathcal{K}_{n}^{(3)} (n=1,2,3,4)$ and $\mathcal{K}_1^{(3)}=\{1,4\}$, $\mathcal{K}_2^{(3)}=\{2,4\}$, $\mathcal{K}_3^{(3)}=\{1,2,4\}$, $\mathcal{K}_4^{(3)}=\{1,3,4\}$. In this situation, we need the following lemma
\begin{lemma}\label{lem:ThmProof}
  Given four nonnegative variables $y_1$, $y_2$, $z_1$, $z_2$ and  $y_1+y_2=z_1+z_2$, we have
  \begin{equation}\label{eq:y12z12}
    \sqrt{y_1}+\sqrt{y_2}\leq \sqrt{z_1}+\sqrt{z_2},
  \end{equation}
  if and only if $|y_1-y_2|\geq |z_1-z_2|$.
\end{lemma}
{\textbf{Proof}}: Denote $y_1+y_2=z_1+z_2=c_0$. We can easily prove that $y_1 y_2 \leq z_1 z_2$ if and only if $|y_1-c_0/2|\geq |z_1-c_0/2|$. On the other hand, $\sqrt{y_1}+\sqrt{y_2}\leq \sqrt{z_1}+\sqrt{z_2}$ if and only if $y_1 y_2 \leq z_1 z_2$ when $y_1,y_2,z_1,z_2\geq 0$ and $y_1+y_2=z_1+z_2$. Furthermore, we know that $|y_1-c_0/2|\geq |z_1-c_0/2|$ if and only if $|y_1-y_2|\geq |z_1-z_2|$ when $y_1+y_2=z_1+z_2=c_0$. This complete the proof of Lemma~\ref{lem:ThmProof}.

With the conclusion presented in Lemma~\ref{lem:ThmProof}, we can easily prove the constraints are fulfilled in this situation. Taking $\mathcal{K}=\mathcal{K}_2^{(3)}$ as an example, we have
\begin{eqnarray*}
  \sum_{k\in \mathcal{K}_2^{(3)}}\tilde{\beta}_k \tilde{x}_k &=&n_0 \sqrt{\tilde{\beta}_2 + \tilde{\beta}_3 + \tilde{\beta}_4}-n_0\sqrt{\tilde{\beta}_3 + \tilde{\beta}_4} \nonumber \\
  & & + n_0 \sqrt{\tilde{\beta}_4} \leq n_0 \sqrt{\tilde{\beta}_2+ \tilde{\beta}_4}.
\end{eqnarray*}
As defined above, $\tilde{\beta}_k\geq 0 (k=1,2,3,4)$. In the last inequality, we have used Lemma~\ref{lem:ThmProof} with $y_1=\tilde{\beta}_2 +\tilde{\beta}_3 +\tilde{\beta}_4$, $y_2=\tilde{\beta}_4$ and $z_1=\tilde{\beta}_2 +\tilde{\beta}_4$, $z_2=\tilde{\beta}_3 +\tilde{\beta}_4$.

In the last situation, there is only one constraint remained with $\mathcal{K}=\mathcal{K}_1^{(4)}=\{1,3\}$. We need to prove that $\sqrt{\tilde{\beta}_1+ \tilde{\beta}_2+ \tilde{\beta}_3 + \tilde{\beta}_4}-\sqrt{\tilde{\beta}_2 + \tilde{\beta}_3 + \tilde{\beta}_4} + \sqrt{\tilde{\beta}_3 + \tilde{\beta}_4}- \sqrt{\tilde{\beta}_4}\leq \sqrt{\tilde{\beta}_1 + \tilde{\beta}_3}$. Equivalently, we need to show that $\sqrt{\tilde{\beta}_1+ \tilde{\beta}_2+ \tilde{\beta}_3 + \tilde{\beta}_4}+ \sqrt{\tilde{\beta}_3 + \tilde{\beta}_4}\leq \sqrt{\tilde{\beta}_2 + \tilde{\beta}_3 + \tilde{\beta}_4} + \sqrt{\tilde{\beta}_1 + \tilde{\beta}_3} + \sqrt{\tilde{\beta}_4}$. We can easily prove this inequality by taking squares of two sides twice with eliminating the same terms on the two sides in each step.

Conclusively, the upper bound $f_{max}$ is reachable with $P_s$ locates in the feasible region of the LP problem. That is to say, $f_{max}$ is really the maximum value of $f(\tilde{x}_k)$ with $\tilde{x}_k=\tilde{x}_k^{*}$. Similarly, we know that the minimum value of $f(x_k)$ is $f_{min}=f(-\tilde{x}_k^{*})=-f_{max}$. This complete the proof of Theorem~\ref{thm:LPSolver} with $K=4$. The conclusions with $K=2,3$ can be proved in the same way.


\end{document}